%% file: NHB27.tex
\documentclass[a4paper,onecolumn,10pt]{article}
\usepackage{amssymb,amsfonts,amsmath}
\usepackage{float}
\usepackage{color}
\usepackage{widetext}
\usepackage[font=footnotesize]{caption}
\usepackage{graphicx}

\usepackage{dblfloatfix}

\usepackage{caption}

\usepackage{bbm}
\usepackage[colaction]{multicol}

\usepackage[default]{lato}
\usepackage[T1]{fontenc}

\usepackage{parskip}
\usepackage[top=0.8in, bottom=1.5in, left=0.5in, right=0.5in]{geometry}
\usepackage{cleveref}

\usepackage{titlesec}
\titleformat{\chapter}[display]
  {\normalfont\sffamily\huge\bfseries\color{blue}}
  {\chaptertitlename\ \thechapter}{20pt}{\Huge}
\titleformat{\section}
  {\large}
  {\thesection}{1em}{}
\titleformat{\subsection}
  {\normalsize}
  {\thesubsection}{1em}{}
  
\usepackage{fancyhdr}
\usepackage[super,comma,compress]{natbib}

\begin{document}

{\huge Productive Ecosystems and the Arrow of Development} 
\\
\\
{Neave O'Clery\textsuperscript{1,3}, Muhammed A. Y\i ld\i r\i m\textsuperscript{2,3} \& Ricardo Hausmann\textsuperscript{3}}
\\
\\
{\footnotesize 1 Mathematical Institute, University of Oxford, UK.}\\
{\footnotesize 2 Ko\c{c} University, Turkey.}\\
{\footnotesize 3 Center for International Development, Harvard University, USA.}

\begin{abstract}

Economic growth is often associated with diversification of economic activities. Making a product in a country is dependent on having, and acquiring, the capabilities needed to make the product, making the process path-dependent. Taking an agnostic view on the identity of the capabilities, we derive a probabilistic model to describe the directed dynamic process of capability accumulation and product diversification of countries. Using international trade data, we identify the set of pre-existing products that enables a product to be exported competitively. This is the \emph{ecosystem} of the product. We construct a directed network of products, the Eco-Space, where the edge weight is an estimate of capability overlap between products. We uncover transition products and a core-periphery structure, and show that low and middle-income countries move out of transition products and into the core of the network over time. Finally, we show that our network model is predictive of product appearances.

\end{abstract}

\begin{multicols}{2}

There is strong evidence that as countries experience economic growth, they change what they do and undergo structural transformation via diversification of their economic activities \cite{imbs2003stages, Atlas}. Emergence of a particular industry in a country depends on availability of different combinations of capabilities, including various factors like capital, labour, and productive knowledge \cite{Hausmann2007, HidalgoHausmann2009, Neffke2013SkillRelatedness}. From this viewpoint, countries grow as they acquire productive knowledge and/or `capabilities', and learn to combine these complementary capabilities in order to move into new economic activities. Hence, industrialisation is a path dependent process, whereby the appearance of new industries and economic activities is conditional on having or acquiring the relevant capabilities and know-how \cite{Hausmann2007, HidalgoHausmann2009, Neffke2013SkillRelatedness, NelsonWinter1982, Frenken2007related, ellison2010causes}. 

Drawing up an exhaustive list of capabilities and/or the productive knowledge required for an industry is challenging. For instance, for a country to develop the fresh cut flower industry, it requires capabilities such as cold storage facilities, airports, irrigation systems, suitable climate, efficient customs, a good business environment as well as knowledge embedded in its farmers, botanic experts, engineers, logistic specialists, marketing professionals, bureaucrats and business executives to name but a few. This list is by no means exhaustive and the listed components might not be independent of each other. Since these capabilities are difficult to observe and measure, we take an agnostic view about their identities and seek to quantify their existence drawing inspiration from biology and, in particular, the study of genetics. In genetics, observed phenotypes are the result of genotypes encoded in genetic material. Mendel, in his landmark study, recorded the phenotypes present in successive generations of peas without directly observing the underlying genes and DNA structure. Hence, valuable information can be gathered by observing the phenotypic traits of individuals when the underlying genetic structure is unknown. Furthermore, by observing which phenotypic traits often co-occur in individuals, or which traits often follow each other, we can uncover genetic relationships or distances\cite{hidalgo2009dynamic}. The genetic distance between phenotypes is relevant, for example, to inferring relationships between diseases\cite{goh2007human}. 

Here we take an agnostic view on the identity of the capabilities and we derive a probabilistic model to describe the directed dynamic process of capability accumulation and product diversification of countries. We  use the presence and appearance of industries in countries (phenotypes) to infer capability and know-how-based (genotypic) relationships between industries. Using our genetics-inspired industry capability distance, and modelling industrial diversification as a process by which countries accumulate capabilities and move into new industries that share existing capabilities, we can predict the emergence of new industries, and hence, decipher the arrow of development.

%%%%%%%%%%%%%%%%%%%%%%%%%%%%%%%%%%%%%%%%%%%%%%%%%%
\begin{table*}[t!]
Box 1| Toy example \\
\vspace*{0.2cm}
\fbox{
\begin{minipage}{8.9cm}
\begin{small}
\vspace*{0.1cm}
A simple toy example illustrates the main concept behind the calculation of the ecosystem for a product. 

\vspace*{0.2cm}
\includegraphics[width=8.9cm]{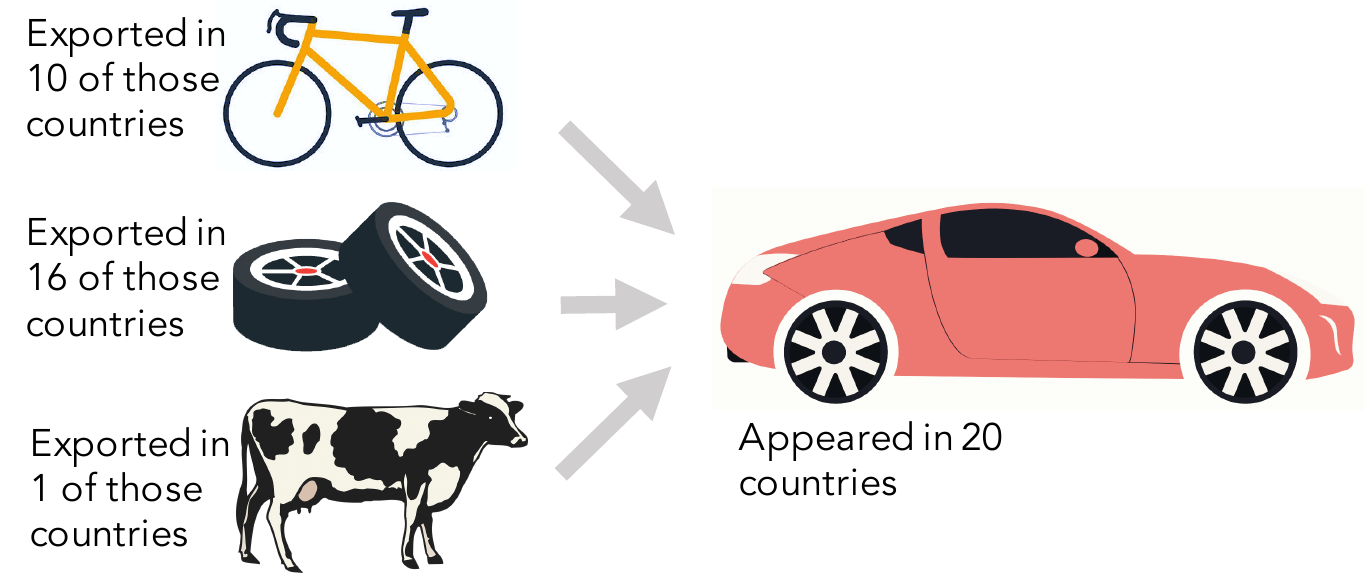}
%\vspace*{0.2cm}

We compute the ecosystem entries for product \emph{car}. Here $N=100$ is the total number of countries and $N_A=20$ is the total number of appearances of cars at $t_1$. We observe $N_P$, the total number of presences of each product at $t_0$, and $N_{AP}$, the number of presences of the product in countries where cars appeared at $t_1$. 

\vspace*{0.1cm}
\end{small}
\end{minipage}
\begin{minipage}{0.2cm}
\end{minipage}
\begin{minipage}{8.9cm}
\begin{small}
%\vspace{-0.05cm}
\includegraphics[width=8.9cm]{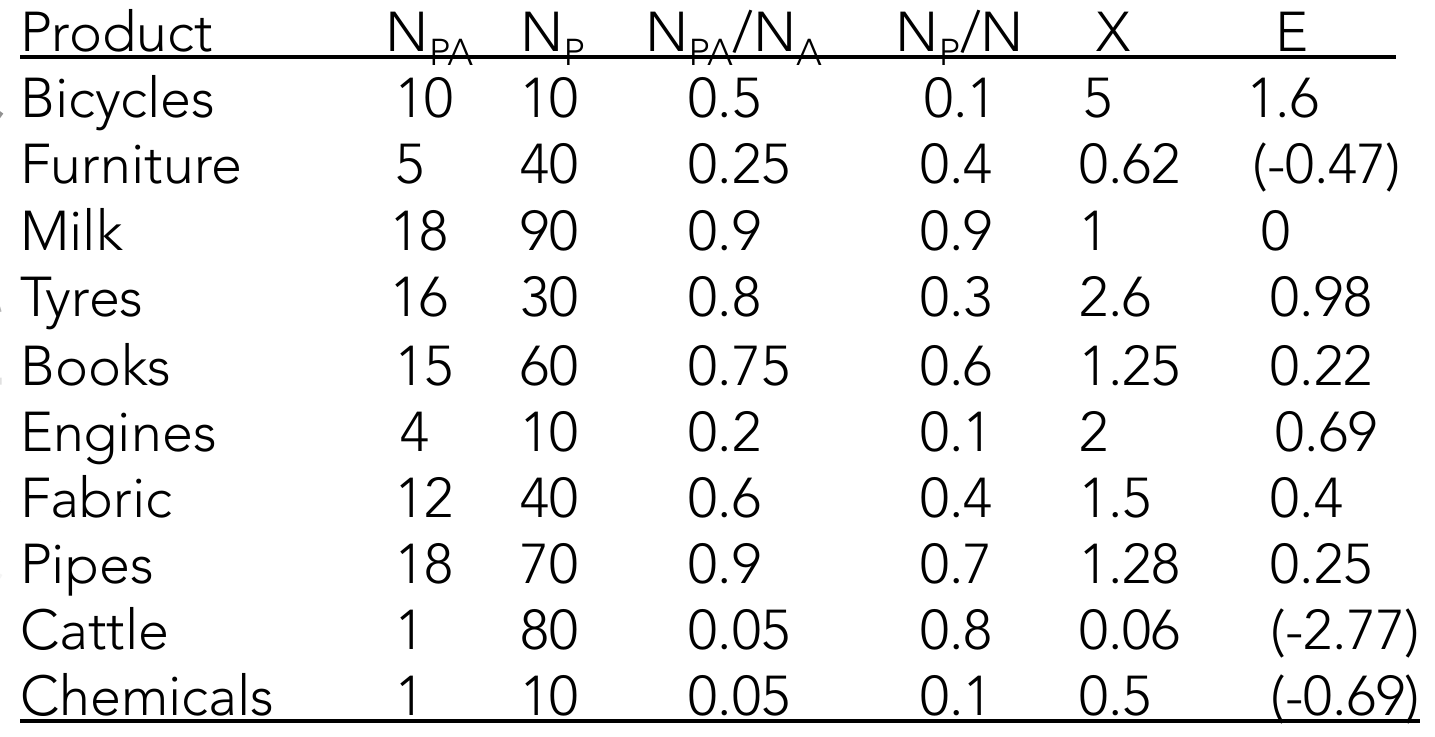}

In this case engines, bicycles and tyres were the most 'over-produced' in the earlier period by countries who later had an appearance of car as shown by $N_{PA}/N_A$. $LR$ is the likelihood ratio of this compared to what would be expected ($N_P/N$). Finally, $E=\log(LR)$ contains the ecosystem entries. Note: entries less than 0 correspond to a ratio $LR$ less than one. 

\end{small}
\end{minipage}
}
\label{model}
\end{table*}
%%%%%%%%%%%%%%%%%%%%%%%%%%%%%%%%%%%%%%%%%%%%%%%%%%

The genetic and phenotypic perspective can be used to reinterpret a number of well-established models of economics as well. For instance, standard trade theories first proposed by Ricardo\cite{ricardo1819principles}, and Hecksher and Ohlin\cite{flam1991heckscher} take complementary perspectives which can be thought as phenotypic and genotypic stances, respectively, to explain trade patterns between countries. For example, a recent and celebrated version of the Ricardian model developed by Eaton and Kortum\cite{eaton2002technology} proposes that technological differences across countries, and the relative evolution of productivity across exports, determines the pattern of production in the world. These authors do not seek to uncover the causes behind the observed pattern, hence taking an implicitly phenotypic view of the international trade. On the other hand, the Hecksher-Ohlin model ties trade patterns to factor differences between countries, and proposes that the relative abundance of factors (labour, capital etc.) shapes the production choices of a country. This model takes a genetic perspective, however, quickly becomes intractable for large numbers of factors and products, constraining detailed insights into diversification processes. 

Turning to models of structural transformation and diversification, understanding these processes at a detailed level has been of keen interest for policymakers and practitioners. However, analytical intractability and measurement problems force economists to often focus on few core productive factors such as capital, labour, human capital and institutions \cite{solow1956contribution, mankiw2014principles} and technological differences \cite{Romer1990,aghion1990model, aghion1998endogenous}, usually taking a genetic perspective. But these models struggle to adequately describe structural transformation at a disaggregate level. Here, we can exploit the fact that we can observe and measure the phenotypes, namely the presence of industries in countries, and propose a phenotypic approach to modelling the process underlying structural transformation at a detailed level.

To date, two coupled but distinct modelling approaches have emerged aiming to describe the path-dependent process of diversification using capabilities and productive knowledge using a phenotypic view. The first is focused on empirically estimating the 'number' of complementary capabilities, or complexity, needed to make a product (or present in a technology or place)\cite{HidalgoHausmann2009, balland2017geography}. While a variety of approaches have been proposed, the foundational method to estimate product complexity\cite{HidalgoHausmann2009, BustosEtal2012Nestedness} uses information on which countries make what products to infer capability requirements under the assumption that complex products can only be made by countries which have many capabilities, and hence, also make many other products. It has been shown that the aggregate complexity level of a country is a strong predictor of its future income growth compared to standard variables often associated with country sophistication such as education and quality of government. A second class of models seeks to map the path dependent dynamical process by which countries move into new products\cite{Hausmann2007}. These are connected, both theoretically and methodologically, to the study of regional and urban industrial diversification\cite{Frenken2007related, ellison2010causes, Neffke2013SkillRelatedness}, and are based on the assumption that countries will move into products similar to their current export (capability) basket. At the forefront of these models, the {\it The Product Space}\cite{Hausmann2007} is a network of products with edges based on cross-sectional export data. Under the assumption that a product pair requiring similar capabilities will be co-exported by many countries, the (cross-sectional) co-export probability of any two products is assumed to be related to the capability overlap. The location of a country in this network (e.g., its subgraph of existing products) determines its future diversification potential. Countries in denser parts of the network have more options, while those on the periphery share capabilities with few other products. The ability of this network model, and others like it, to generate detailed metrics related to diversification processes has propelled the field into development and industrial policy-making at the global, national and regional level\cite{Atlas, boschma2014constructing}.

Yet, these dual modelling approaches, capturing slightly different elements of the same underlying process, have not been unified to date. Additionally, they do not address the temporal aspect of the diversification process as a result of capability accumulation directly. Furthermore, they omit a large amount of available information on the patterns of diversification observed over the past couple of decades worldwide. Here we seek to develop a unified model, which is theoretically grounded with the path dependent accumulation of capabilities and products, and utilises the available data for international export diversification since the mid 1980s.

%%%%%%%%%%%%%%%%%%%%%%%%%%%%%%%%%%%%%%%%%%%%%%%%%%
\begin{figure*}[t!]
\centering
\includegraphics[width=18.5cm]{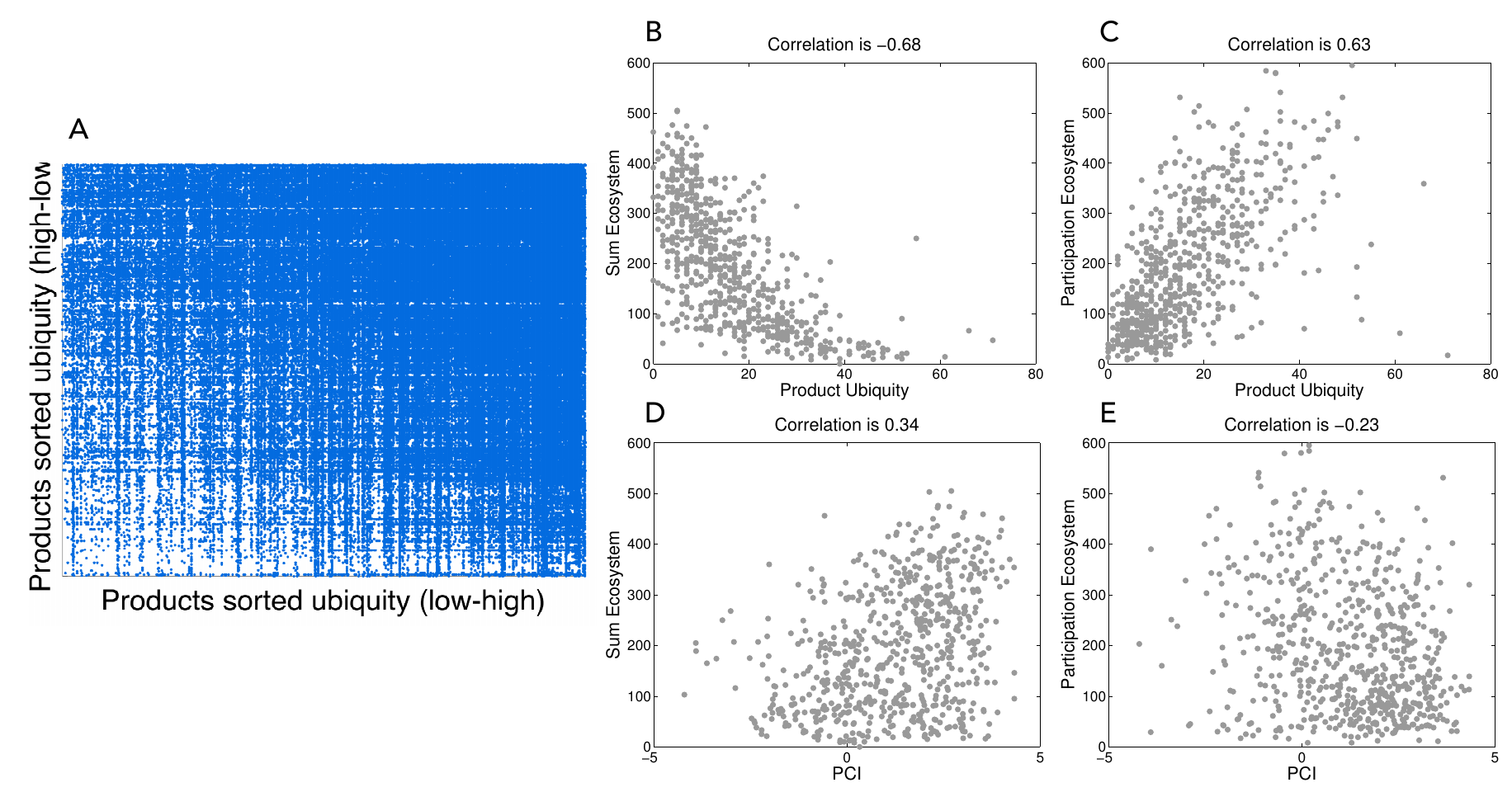}
\caption{(A) We create a single composite ecosystem matrix using data from 1984-2008. Rows contain the ecosystem entries for each product. Products are sorted by ubiquity or rarity on both axis. We can see that rare products require many inputs (top rows are densely filled), and common products are inputs to many products (right-hand columns are densely filled). 
(B-C) We confirm these observations by showing a negative relationship between the row-sum of positive entries vs product ubiquity, and a positive relationship between the column-sum of positive entries for each column vs product ubiquity. 
(D-E) We also consider an alternative measure of product complexity \cite{Hausmann2007, HidalgoHausmann2009}, the product complexity index (PCI - computed in 1984). We observe a similar pattern: large ecosystems are associated with complex products, and less complex products contribute to many ecosystems.}
\label{fig1}
\end{figure*}
%%%%%%%%%%%%%%%%%%%%%%%%%%%%%%%%%%%%%%%%%%%%%%%%%%

Building on Hausmann and Hidalgo\cite{HausmannHidalgo2011}, who developed a capability-based Leontief-like production function, we propose a model to describe the pattern of product appearances within and across countries based on capability accumulation. Within this framework, a country will jump to a new product with probability decreasing with the number of missing capabilities. We measure the capabilities possessed by a country by looking at the capabilities of the products it currently produces. The ability of a country to diversify is, hence, dependent on its current product basket. Countries with many existing products will have few missing capabilities, and many options for diversification. Hence, the pattern of product appearances contains information about the underlying capability overlap between products. We derive the relationship between the probability of a product presence (say product $i$) given the subsequent appearance of product $j$, and use this to infer the extent of capability overlap between the product pair $i$ and $j$. The \emph{ecosystem} of a product $i$ is then the overlap of product $i$ with all other products $j$. We empirically estimate this capability overlap using product presences and appearances in international export data from 1984 to 2015. 

In order to explore path dependent diversification processes, we construct a weighted directed network, the {\it Eco Space}. The direction of the edges connecting nodes (products) represents export precedence, and the edge weight is given by our estimate of capability overlap. We analyse a range of network characteristics including betweenness centrality, a measure of the number of shortest paths that transition through a node. Such nodes exhibit both high in- and out-degree - they are transition products typically produced by low and middle income countries as they move into more sophisticated products. Nodes with high in-degree, on the other hand, are typically complex products, requiring many inputs. We observe a core-periphery structure \cite{borgatti2000models} whereby a densely connected core composed of high in-degree products is surrounded by a less connected periphery. Such a structure has previously been observed in world trade networks \cite{smith1992structure,della2013profiling}. We show that low and middle-income countries move over time out of transition products and into the core. 

Finally, using an out-of-sample approach, we show that our model (empirically estimated from export data for the period 1984-2008) is informative in predicting the emergence of new products in the exports of countries for the period 2009-2015. We can interpret this result as suggesting that a country with an export basket proximate (in terms of capability gap) to a particular product is more likely to competitively export that product in the future. This model compares favourably in comparison to the Product Space\cite{Hausmann2007} in terms of the prediction of export appearances. Importantly, together these measures provide significant predictive power, suggesting that this model contains complementary information, which we hypothesise includes additional information on capability accumulation not captured by the cross-sectional Product Space.

%%%%%%%%%%%%%%%%%%%%%%%%%%%%%%%%%%%%%%%%%%%%%%%%%%
\setlength{\unitlength}{1cm}
\begin{table*}[t!]
\begin{picture}(20,0.001)
\put(1.6,-0.3){\bf A}
\put(9.1,-0.3){\bf B}
\end{picture}
\begin{center}
\begin{footnotesize}
\begin{tabular}{|l|}
LARGEST ECOSYSTEM PRODUCTS \\
\\
\input{indeg_new.tex}
\end{tabular}
\hspace{0.5cm}
\begin{tabular}{|l|}
TOP CONTRIBUTING PRODUCTS \\
\\
\input{outdeg_new.tex}
\end{tabular}
\end{footnotesize}
\end{center}
\caption{(A) Here we show the top 15 products in terms of positive ecosystem entries (e.g., the number of positive entries on the rows of $\hat{E}$). (B) The top 15 products in terms of contribution to product ecosystems (e.g., the number of positive entries on the columns of $\hat{E}$). In the first case we observe a range of sophisticated products including engines, chemicals, equipment and vehicles. In the second case, overall we have less complex products including food, textiles, metals and basic chemicals. \label{tab1}}
\end{table*}
%%%%%%%%%%%%%%%%%%%%%%%%%%%%%%%%%%%%%%%%%%%%%%%%%%

In related work, Zaccaria \emph{et al}\cite{zaccaria2014taxonomy}, similarly inspired by the capability-based model, created a taxonomy of products based on the excess conditional probability of producing a product in the presence of other products. Like the Product Space, this method computes the probabilities using cross-sectional export data. By selecting the maximum among the excess probabilities, they generate a product hierarchy tree and use this to model the dynamics of product diversification of countries. Although the underlying justification is based on the capability model, neither the Product Space nor the taxonomy method explicitly take capabilities into account in their modelling approach. 

Our framework can be potentially applied to a range of other settings where path-dependent diversification occurs. The first obvious extension is to the regional setting where firms/industries need specific locally available capabilities to flourish. This will result in a path-dependent process of diversification, and underlies some of the forces behind industrial cluster formation\cite{porter1998clusters} and urban agglomeration\cite{ellison2010causes}. In a second extension, technology adoptions by countries\cite{comin2010exploration} also follow a path-dependent process: many technologies require other technologies to be present in advance in a country. In biology, from where we borrow the term ecosystem, organisms require the presence of other animals or plants to populate a location, and, hence, this mechanism also leads to path-dependent dynamics. This process is intimately linked to observed nested structure emphasised in the ecology (and economics) literature\cite{bascompte2003nested, saavedra2009simple, saavedra2011strong, BustosEtal2012Nestedness}.

%%%%%%%%%%%%%%%%%%%%%%%%%%%%%%%%%%%%%%%%%%%%%%%%%%
\section*{Results}
%%%%%%%%%%%%%%%%%%%%%%%%%%%%%%%%%%%%%%%%%%%%%%%%%%
\subsection*{Productive Ecosystems \label{simple_model}}
%%%%%%%%%%%%%%%%%%%%%%%%%%%%%%%%%%%%%%%%%%%%%%%%%%

In order to model the process of product diversification via capability accumulation, we build on Hausmann and Hidalgo\cite{HausmannHidalgo2011}. According to this Leontief-like model, products require a large number of capabilities in order to be made, and countries can only make a product if they possess all the required capabilities. We denote the vector of capabilities of a product $i$, $p_i \in \{0, 1\}^m$ where  $m$ is the number of capabilities and $p_{ik}=1$ if product $i$ requires capability $k$. Analogously, the capability vector $c_n \in \{0, 1\}^m$ encodes the capabilities present in country $n$. 

We assume that country $n$ will start making a product $i$ at a future time $t_1$, which it does not currently make, with a probability that decreases with the number of capabilities that are not present in the country at some initial time $t_0$. Formally, we are concerned with the capability 'gap' between the capability vector of the country and capability requirement vector of the product: $| p_i | - c_n \cdot p_i$. The probability that country $n$ will start making product $i$ decreases as size of this gap increases. Following Hausmann and Hidalgo\cite{HausmannHidalgo2011}, we can assume that the probability of acquiring a capability is binomial with mean $q$. Hence, 
\begin{equation}
\label{capgap}
P(J_{n,i} = 1) = q ^ {| p_i | - c_n \cdot p_i }
\end{equation}
where $J_{n,i}=1$ if product $i$ appears in country $n$ at time $t_1$, and 0 otherwise. 

We show in the \emph{Methods} section that, if we assume that the probability of a country having each capability is $w$, we can express the the capability overlap between $i$ and $j$ as 
\begin{equation}
E_{i,j}=\log\left(\frac{P(M_{n,j} = 1 |J_{n,i} = 1)}{P(M_{n,j} = 1)}\right)  = -| p_i^j | (1-w)\log(q)
\label{eco}
\end{equation}
where $M_{n,i}=1$ if product $i$ is present in country $n$ at $t_0$, and $E$ is the ecosystem matrix. Therefore, the probability that the product $j$ is already produced in a country, given the country started making the product $i$, increases with the overlap between the capability requirements of these two products, captured by $|p_i^j|$ up to a constant multiplicative factor. 

We refer to the vector $E_i=\{E_{i,j}\}_{j=1,...,n}$ as the ecosystem of a product $i$. This captures the extent of capability overlap between product $i$ and all other products $j$, and is calculated based on the probability that product $j$ was already present when product $i$ appeared. In the \emph{Methods} section we outline how we empirically estimate the ecosystem matrix using product presences and appearances (based on revealed comparative advantage\cite{balassa1965trade}) in international trade data. In order to study long term diversification trends, we create a single composite ecosystem matrix $\hat{E}$ using data from 1984-2008 for 674 4-digit SITC products.\footnote{We construct the ecosystem using data from 1984-2008. Below we use an out-of-sample approach to predict product appearances for the period 2009-2015.}$^,$\footnote{Negative values are set to 0 in this matrix. This corresponds to a ratio $P(M_{n,j} = 1 |J_{n,i} = 1)/P(M_{n,j} = 1)$ less than 1.} A toy example illustrating the method is shown in Box 1. 

%%%%%%%%%%%%%%%%%%%%%%%%%%%%%%%%%%%%%%%%%%%%%%%%%%
\begin{figure*}[t!]
\centering
\includegraphics[width=18.5cm]{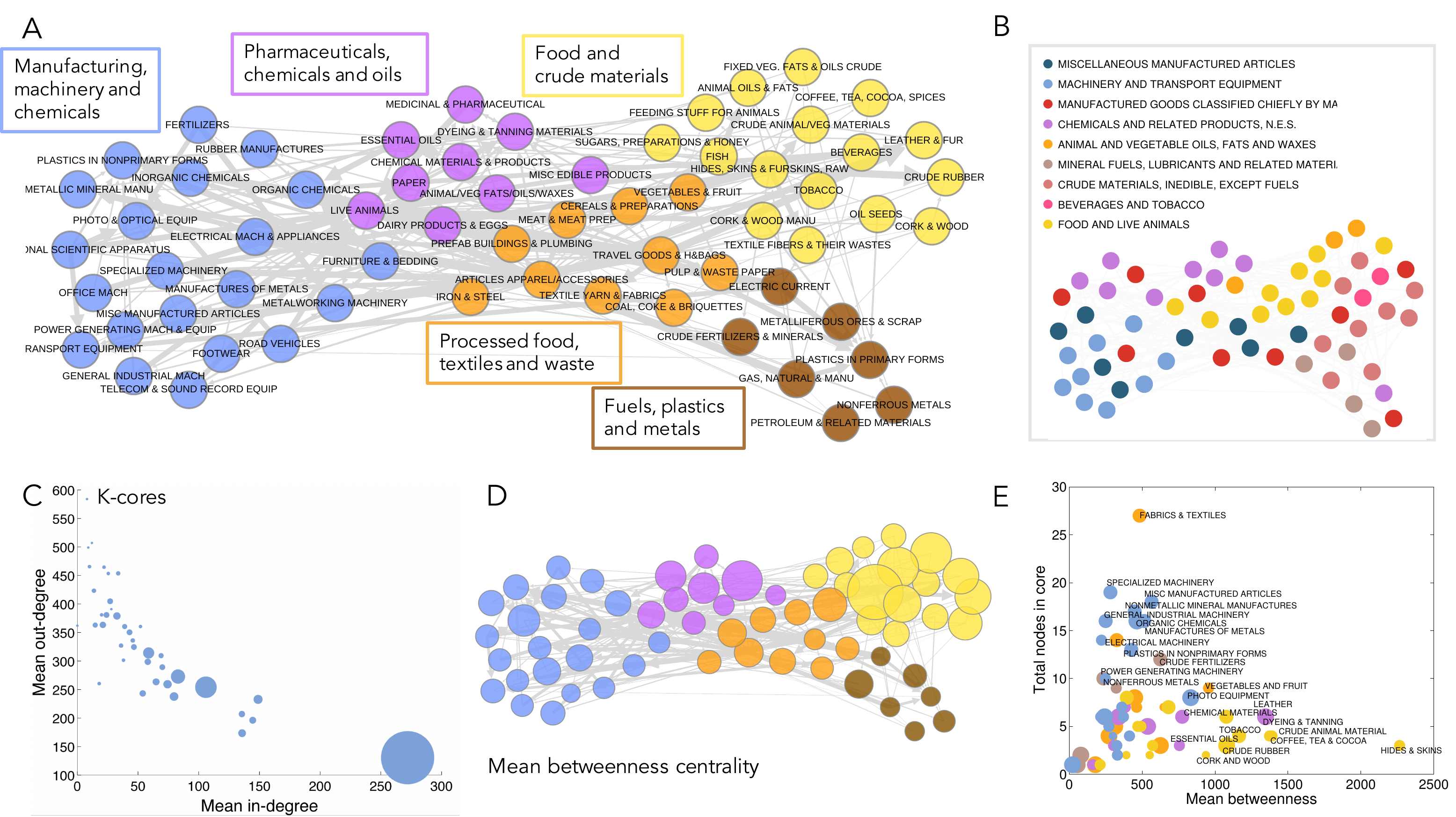}
\caption{
(A-B) 
For visualisation purposes, we show the directional relationship between 2-digit products. The edges are the mean edges weights between 2-digit node groupings in the 4-digit product network (not shown). Nodes are coloured by node community. On the right-hand-side, we show the node colouring by 1-digit product sector.
(C)
We show the mean in-degree vs out-degree for nodes in each k-core of the 4-digit network. The marker for each core is sized proportional to the number of nodes in the core. We can clearly see what there is a single densely connected large core, surrounded by smaller layers, containing mainly high in-degree (large ecosystem) nodes. 
(D)
We compute the betweenness centrality of each node (in the 4-digit network), a measure of the number of times a shortest path between any two nodes traverses the node, and visualise the mean for each 2-digit sector. These 'transition' products have high in and out degree - and include mainly processed foods, beverages, non-fuel crude materials, basic chemicals and manufacturing such as textiles. These are primarily located outside the core on the left-hand side of the network.
(E) 
Finally, we show the mean betweenness centrality vs total nodes in core for 2-digit sectors (markers are coloured by community, and sized by the share of nodes in the core). We observe distinct behaviour for machinery (blue markers tend to have many nodes in the core) and processed food/materials and basic chemicals/plastics (yellow and pink tend to be transition sectors).  
%While manufacturing, chemicals and machinery dominate the core, food and raw materials act as transition products. 
}
\label{network}
\end{figure*}
%%%%%%%%%%%%%%%%%%%%%%%%%%%%%%%%%%%%%%%%%%%%%%%%%%

Products with large ecosystems (e.g, many non-zero entries in their ecosystem vector) share capabilities with many other products. These are sophisticated products, likely made by few countries. Drawing on previous work\cite{BustosEtal2012Nestedness}, we investigate whether rare products have larger ecosystems, and common products tend to be part of many ecosystems. 
Figure \ref{fig1} (A) presents a visual representation of the entries in matrix $\hat{E}$. Products are sorted by the number of countries making them (ubiquity) on both axis. Hence, products occupying the top rows, and left-most columns, are rare. 
We observe that rare products require many inputs (top rows are densely filled), and common products functions as stepping stones to many products (right-hand columns are densely filled). 
Set $X=\hat{E}>0$ (e.g., an indicator matrix for the positive entries of $\hat{E}$). We confirm these observations by showing a negative relationship between the row-sum of $X$ and product ubiquity, and a positive relationship between the column-sum of $X$ and product ubiquity in Figure \ref{fig1} (B) and (C). This behaviour is consistent with the nested pattern observed in cross-sectional data for product presences by Bustos \emph{et al} \cite{BustosEtal2012Nestedness}.

As discussed above, a range of approaches have been proposed to quantify the complexity of a product. The product complexity index (PCI)\cite{HidalgoHausmann2009}, a widely-used measure of product sophistication, is derived from export data based on the hypothesis that rare or 'complex' products are only made by few countries that possess many capabilities. Echoing the nested behaviour discussed above, higher complexity products are mostly associated with (rare) developed countries while lower complexity products are produced in countries at all levels of development. Figure \ref{fig1} (D) and (E) show more complex products are associated with larger ecosystems, and less complex products contribute to many ecosystems.

Table \ref{tab1} shows the top 15 products in terms of positive ecosystem entries (e.g., the row-sum of $X$), and the top 15 products in terms of contribution to product ecosystems (e.g., the column-sum of $X$). In the first case we observe a range of sophisticated products including machinery engines, chemicals, equipment and vehicles. In the second case, overall we have less complex products including food, textiles, simple machinery and basic chemicals.

%%%%%%%%%%%%%%%%%%%%%%%%%%%%%%%%%%%%%%%%%%%%%%%%%%
\begin{figure*}[t!]
\includegraphics[width=18.5cm]{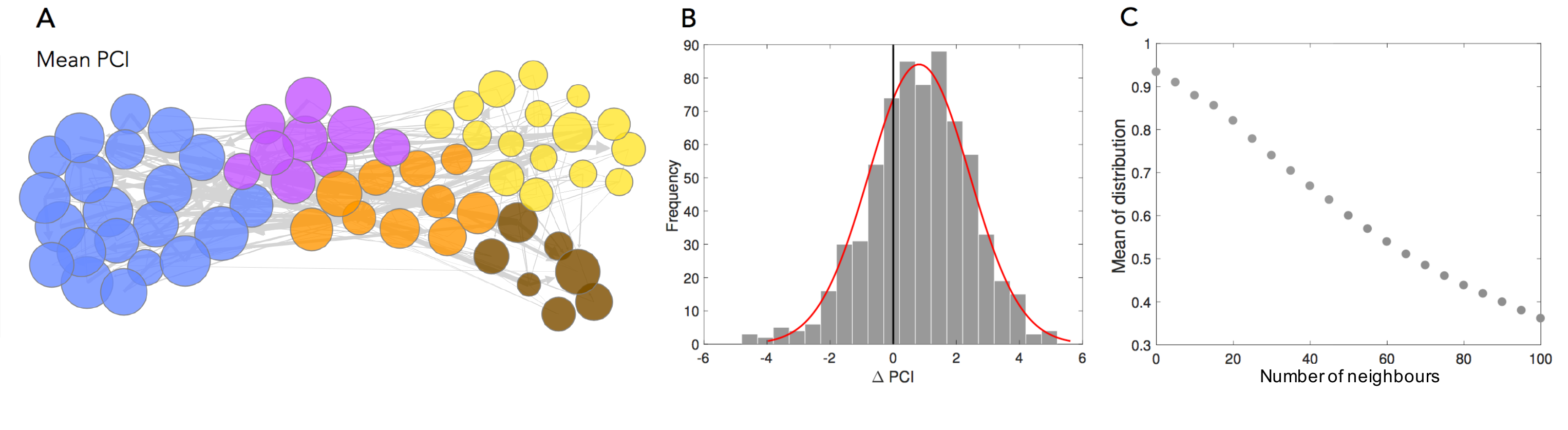}
\caption{(A) Here we show the mean PCI of products in 2-digit sector. High complexity nodes, as we have seen, correlate with products exhibiting a large ecosystem. (B) We are interested to see whether the directionality of edges moves from lower to higher PCI products. For each product, we compute the PCI of the product minus the mean PCI of its top $x=10$ incoming neighbours. The histogram shows a clear bias towards positive values - the PCI of the product is higher than its incoming neighbours. (C) By looking at the mean of the distribution across a range of $x$, we find, as expected, this mean moves towards zero as we increase the number of neighbours. \label{pci}}
\label{fig:pci}
\end{figure*}
%%%%%%%%%%%%%%%%%%%%%%%%%%%%%%%%%%%%%%%%%%%%%%%%%%

\subsection*{The Arrow of Development}
 
Countries diversify into new products that are similar (in terms of required capabilities) to what they currently produce. In order to model this process, we construct a network of products. Directed edges connect the products: there is an arrow from node $j$ to node $i$ if product $j$ is in the ecosystem of product $i$, i.e., $j$ tends to be produced before $i$ appears. The weight of the edge is an estimate of capability overlap between $i$ and $j$ as determined by the corresponding ecosystem entry. 

We can ask questions such as: do we observe clusters of products sharing many capabilities? Which products are most likely to be part of a development path? How do countries diversify in this network?

Formally, we consider a network with $n$ nodes (or vertices). The structure of a network is represented by the adjacency matrix $A \in \mathcal{R}^{n \times n}$ where entries $A_{ij}$ correspond to the weight of the directed edge from node $i$ to node $j$. In this case, $A=\hat{E}^T$ is the adjacency matrix for the {\emph Eco Space}. Using this adjacency matrix we can compute a host of network metrics, including, for example, the in-degree for each node $i$, $d_i=\sum_j X_{ji}$, and the in-strength $s_i=\sum_j A_{ji}$\footnote{The in-degree is exactly the sum of the positive ecosystem entries shown Figure \ref{fig1}. Similarly, the node out-degree is the ecosystem contribution.}.
 
For visualisation purposes, we compute a reduced version of this network, calculating the mean edge weight between products in 2-digit sectors (there are 63 2-digit product sectors). Figure \ref{network} (A) illustrates the directional relationship between sectors, showing mean edge weights over a threshold of 0.15 (this includes about the top 12\% of edges)\footnote{The software programme Gephi \cite{bastian2009gephi} has been used to generate the network layout (using the automated Force Atlas algorithm).}. 
We can examine clustering of sectors with shared capabilities in this network via community detection. While there are a variety of approaches, here we apply the Stability algorithm\cite{Delvenne2010}, a multi-scale generalisation of the well-known modularity algorithm\cite{f54}. This method is based on the dynamics of a random walker on a network - the walker gets trapped in regions of high connectivity to unearth community structure. 
We observe clear groupings, with food/animals/crude materials dominating the yellow community on the right-hand-side. As we move to the left we observe clusters of processed goods (orange) and fuels (brown), before moving into basic chemicals (pink) and advanced chemicals/manufacturing (blue). Notably, this node partition based on clustering of empirical development paths is relatively consistent with the official 1-digit sector classification shown in Figure \ref{network} (B). 

We now analyse in more detail the base 4-digit network (not shown). We can explore the overall structure of the network by applying a k-core algorithm \cite{borgatti2000models}, typically used to extract nested network layers. A k-core is a maximal connected subgraph in which all nodes have degree at least k. Hence, as k increases we get a more densely connected subgraph. A core-periphery structure such as this has been observed for other economic networks such as trade networks \cite{smith1992structure,della2013profiling}, where a small dense core, composed of countries with a very high volume of trade, induces stability in global trade dynamics. 
In Figure \ref{network} (B) we show the mean in-degree vs mean out-degree for each k-core. The marker for each core is sized proportional to the number of nodes in the core. We can clearly see what there is a single large core (far right), surrounded by smaller layers. In this case, nodes in the largest core tend to have high in-degree - e.g., they require many inputs. Hence the core of this network is composed of sophisticated products requiring many overlapping capabilities. We expect that, as shown below, only advanced economies possessing many capabilities will be able to reach the core. 

%%%%%%%%%%%%%%%%%%%%%%%%%%%%%%%%%%%%%%%%%%%%%%%%%%
\begin{figure*}[t!]
\includegraphics[width=18.5cm]{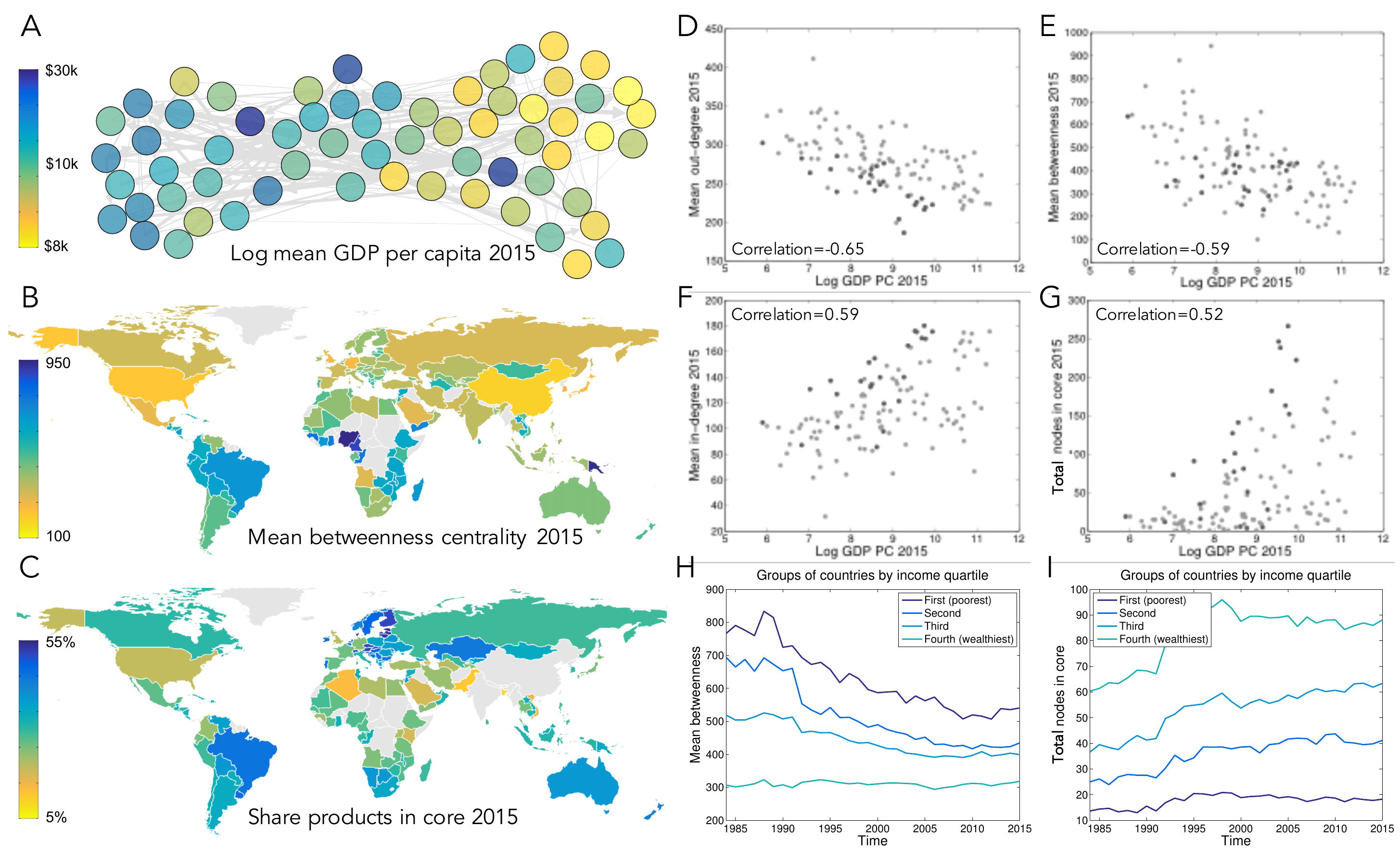}
\caption{
(A-C) 
First, we show the mean GDP per capita for countries with products in each 2-digit sector. As expected, the average GDP per capita falls as we move towards the right of the network dominated by food, animals and crude materials. Next, we show maps with countries shaded by mean betweenness centrality of their products, and total products in the core, in 2015.
(D-G)
The poorest countries dominate products with a large out-degree (contribute to many ecosystems), and transition products with high betweenness centrality. On the other hand, wealthy countries have highest mean in-degree (products with the largest ecosystems), and most products in the core (all data for 2015). 
(H-I) 
Over time, from 1984-2015, we observe that poor countries moved out of transition products with high betweenness centrality, and middle and high income countries increased their share of products in the core.}
\label{countries}
\end{figure*}
%%%%%%%%%%%%%%%%%%%%%%%%%%%%%%%%%%%%%%%%%%%%%%%%%%%

We can also extract information about intermediate steps. We compute the betweenness centrality of each node, a measure of the number of times a shortest path between any two nodes traverses the node. These 'transition' products tend to have high in and out degree - they are stepping stones. Figure \ref{network} (D) and (E) show that this measure is largest for processed food, beverages, non-fuel crude materials, basic chemicals and manufacturing such as textiles. 
Consistent with the capability-based view of economic growth, whereby oil is a natural endowment that rarely promotes the local knowledge acquisition needed for economic growth (related to the so-called 'resource curse' \cite{sachs1995natural}), fuel is not a stepping stone product. Countries that make high-betweenness products are expected to be primed to move into new, more sophisticated products. 
As is evident from Figure \ref{network} (E), we note that sectors tend to either be dominated by transition products or core products (or neither) - with virtually no sectors exhibiting both characteristics. This finding emphasises the role of network structure in the diversification process, and sheds light on product characteristics only available through network analysis. 

If capability accumulation underlies the development process, we expect countries to move from less complex products towards sophisticated products over time. Hence, we expect diversification from low complexity to high complexity products as countries upgrade their complexity level. We look at the edges between nodes of different complexity levels and ask, is it more likely that a directed edge connects a lower complexity node to a higher complexity node? In other words, are the input products within a product's ecosystem less complex than the product itself? 
To reduce issues surrounding endogeneity, we consider the relative PCI of product pairs. Figure \ref{fig:pci} (A) shows the distribution of PCI within the Eco Space (e.g., the mean PCI of products within the 2-digit sectors). As seen above, high PCI products coincide with technologically advanced products (those with a large ecosystem) on the left-hand side. 
We are interested to see whether the directionality of edges moves from lower to higher PCI products. For each node we compute the difference between its own PCI, and the mean PCI of its top $x=10$ incoming neighbours (by subtraction of the latter). Considering the histogram of values, we observe a clear bias towards positive values - the PCI of the product is higher than its incoming nodes. In fact, in this case, 72\% of products have a higher PCI than the mean of their top 10 ecosystem products. Next, we vary the number of neighbours $x$. By looking at the mean of the distribution across a range of $x$, we find, as expected, this mean moves towards zero as we increase the number of neighbours. 

%%%%%%%%%%%%%%%%%%%%%%%%%%%%%%%%%%%%%%%%%%%%%%%%%%
\begin{table*}[t!]
\begin{minipage}{12cm}
\begin{scriptsize}
\input{regs_rpop_eco0_1.tex}
\end{scriptsize}
\end{minipage}
\begin{minipage}{5cm}
\includegraphics[width=6.5cm]{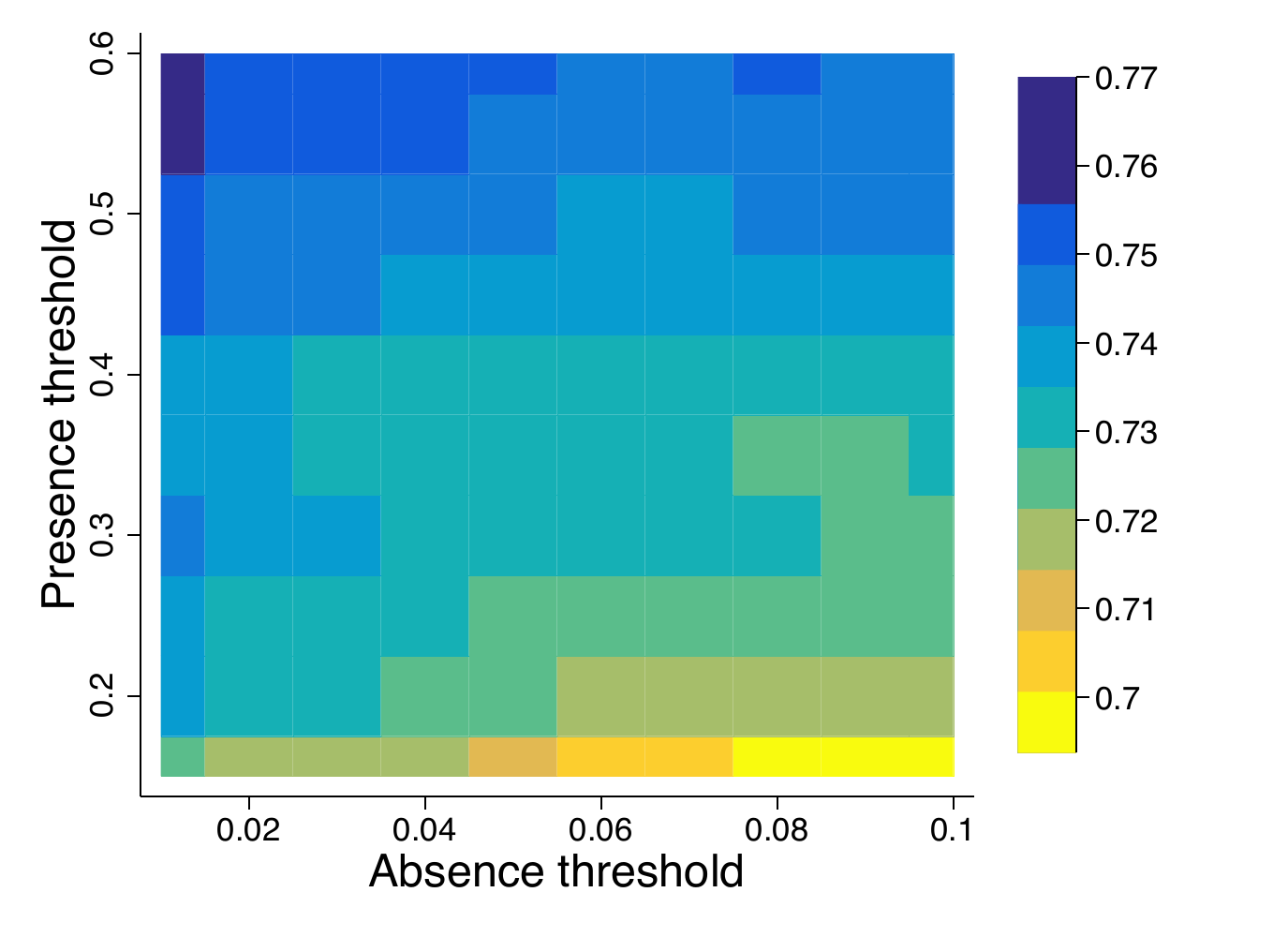}
\end{minipage}
\caption{(A) We seek to predict appearances during the period 2009-2015. For a standard probit model, we see that the ecosystem density metric has predictive power for country-product appearances with AUC=0.736 (column 1), rising to AUC=0.83 when fixed effects are included, and compares favourably to a similar metric based on the Product Space (column 5).  
(B) Heat map for the AUC values for various combinations of parameters $\tau_0$ (absences) and $\tau_1$ (presences). The AUC corresponds to an equivalent regression to column 1 of the table.}
\label{regs}
\end{table*}
%%%%%%%%%%%%%%%%%%%%%%%%%%%%%%%%%%%%%%%%%%%%%%%%%%

How do these product attributes relate to the diversification of countries on this network? 
Figure \ref{countries} (A) shows the mean GDP per capita (2015) for countries\footnote{While the ecosystem matrix was constructed from data including only 109 countries, here we look at the product basket and dynamics for a larger set of 123 countries. The 109 countries were present for the whole period 1984-2008, and have population greater than 1 million people and exports greater than one billion US dollars in a year. The sample increases to 123 if we include countries that were formed during this period.} with products in each 2-digit sector. This measure is similar to the PRODY metric of Hausmann, Hwang and Rodrik (2007)\cite{f14} which associates an average income level to individual products. As expected, the average income falls as we move towards the right of the network dominated by food, animals and crude materials. 
Next we explore node metrics for the products of individual countries. The map in Figure \ref{countries} (B) shows countries shaded by mean betweenness centrality. We observe that darker blue and blue-green countries, those with higher betweenness centrality primed for transition into more complex products, are located in Africa and Latin America. Figure \ref{countries} (C) shows the share of core products, with concentrations in Northern and Central European countries as well as Latin America. 
Overall, as shown in Figures \ref{countries} (D)-(G), the poorest countries dominate products with a large out-degree (contribute to many ecosystems), and transition products with high betweenness centrality. On the other hand, wealthy countries have highest mean in-degree (products with the largest ecosystems), and most products in the core. Figures \ref{countries} (H)-(I) explore the evolution of these metrics over time (1984-2015), dividing countries into four equally-sized income groups. We observe that during this period poor countries moved out of transition products with high betweenness centrality, and middle and high income countries increased their number of products in the core. 

\subsection*{Predicting Product Appearances \label{probit_sect}}

Beyond analysing network properties and diversification paths, we wish to assess whether the model is informative in predicting the appearance of new products, or equivalently the export of new products with comparative advantage, for the set of all countries. 

To predict the likelihood of an appearance of product $i$ in country $c$, as shown in the \emph{Methods} section, we estimate the capability gap in the exponent of Equation \eqref{capgap} via 
\begin{equation}
\label{apps}
d^E_{i,n}=q^{\max_j \hat{E}_{i,j} -\max_{j\in \mathcal{J}_n} \hat{E}_{i,j}}
\end{equation}
where $\mathcal{J}_n$ is the set of products present in country $n$. We note that the ecosystem encoded in matrix $\hat{E}$ was constructed using data from 1984-2008. Product presences in Equation \eqref{apps} are measured in 2009, and seek to predict appearances during the period 2009-2015. 
For a standard probit model, Table \ref{regs} shows that this metric has predictive power for country-product appearances with AUC=0.736, increasing to AUC=0.83 when country and product fixed effects are included.\footnote{The AUC is the \emph{Area Under the Curve} of the Receiver Operating Characteristic (ROC) which plots the rate of true positives of a continuous prediction criterion as a function of the rate of false positives.}

We compare the ability of this metric to predict product appearances with the Product Space density\cite{Hausmann2007,Hausmann2014implied}, an analogous predictive metric based on the structure of the Product Space (see \emph{Methods} for details). We find that our ecosystem based metric outperforms the Product Space density which has AUC=0.664 (column 5, no fixed effects). When both measures are included together, both measures remain significant (both with and without fixed effects) suggesting each contains complementary information. 

Product appearances are dependent on two thresholds: one for product absences ($\tau_0$) and one for product presences ($\tau_1$), see Equations \eqref{defpres} and \eqref{eqn_J} in the \emph{Methods} section. The default values of these, discussed below, are $\tau_0=0.05$ and $\tau_1=0.25$. As we decrease $\tau_0$, we have fewer absences (and hence fewer possible appearances). As we increase $\tau_1$, we also have fewer appearances. 
In order to explore variation in the predictive ability of our model for variation in these parameters, we show a heat map for the AUC values for various combinations of $\tau_0$ and $\tau_1$ corresponding to column 1 of the table. 
We observe, for reasonable combinations $\tau_0$ and $\tau_1$, the base-line (no fixed effects) AUC scores are consistently at least 0.73. 
%
%We observe similar patterns for the value of AUC in each case, indicating robustness of our results to variation in parameters $\tau_0$ and $\tau_1$. 

%%%%%%%%%%%%%%%%%%%%%%%%%%%%%%%%%%%%%%%%%%%%%%%%%%
\section*{Discussion}

Classical growth and trade theory has struggled to reconcile macro variables such as factor endowments with differences in the productive structure and know-how of nations. One approach would be to increase the number of factors measured and write down more detailed production functions to understand the dynamics. A complementary approach might take an agnostic stance towards the identity of the capabilities or factors but focus on the development paths associated with this deeply granular process. In this paper, we took the latter approach and, inspired by early approaches to the study of genetics, we develop a model for product diversification based on capability accumulation, and investigate path dependance in economic development via network analysis.

We propose a new metric, the {\it ecosystem} of a product, which contains information on other products sharing a high-level of capability overlap. Empirically, this is the set of pre-existing products that are typically necessary for a future appearance of that product. Given the temporal nature of this measure, we construct a directed network to describe probable development paths. Exploiting tools from network science, we identify product clusters, transition sectors, and a tightly connected core, governing diffusion on the network. Finally, we show that the model is a good predictor of export diversification, performing favourably compared to the well-known Product Space framework\cite{Hausmann2007}. 

This work contributes to both the theoretical literature on the modelling of capabilities and knowledge accumulation, and more generally the processes underlying economic growth. It is particularly relevant for the literature on economic complexity\cite{HidalgoHausmann2009}, and the on-going search for empirical methods to quantify, measure and validate complexity\cite{Hausmann2007,bettencourt2014professional, gomez2016explaining, balland2017geography}. Similarly, it is embedded in the literature on path dependent diversification\cite{Frenken2007related,Neffke2013SkillRelatedness, Hausmann2007}, including regional dynamics and 'related varieties' similarly derived from an evolutionary or capability-based perspective. Future work could include estimating this model for industry employment or establishment data, which provides additional information on domestic production (and by extension domestic and service capabilities) not contained in export data \cite{Hausmann2014implied,formality}. 

Finally, as doubt in market efficiency has wained, particularly since the 2008-9 financial crisis, industrial policy has enjoyed somewhat of a global resurgence\cite{stiglitz2013introduction}. We hope that the ecosystem metric is helpful to policy-makers seeking to analyse the preparedness of a nation or region to move into a new product, or trying to identify key transition sectors which could open up future opportunities. 
%For example, a city considering automobiles as a future export sector could examine whether it already exports a basket of products typically associated with future success of exporting cars. 

\section*{Methods}

%{\small 
\subsection*{The Model}

Let $M_{n,i}=1$ if product $i$ is present in country $n$, and otherwise 0. Similarly, let  $J_{n,i}=1$ if product $i$ appeared in country $n$, and otherwise 0.

For a product $i$ and a country $n$, $p_i \in \{0, 1\}^m$  is the capability requirement vector of product $i$, and $c_n \in \{0, 1\}^m$ represents the capabilities present in country $n$. Following Hausmann and Hidalgo\cite{HausmannHidalgo2011}, the country $n$ makes the product $i$ if $n$ has all necessary capabilities to make $i$. Formally:
$$ M_{n,i} = 1 \iff |p_i| = c_n . p_i.$$
We assume that the country will jump to the product upon successful completion of all the capabilities its missing to make the product. So, the probability of jump will depend on the capability gap between the country and the product capability vectors:
$$ \Delta_{n,i} =  |p_i| - c_n . p_i.$$ 
We wish to quantify the likelihood of producing product $i$ given the country is already producing product $j$. We can split the capability vector of product $i$ into two parts, one which contains the capabilities overlapping with $j$, and other the non-overlapping capabilities. We write $p_i = p_i^j + \bar{p}_i^j$, where 
\begin{itemize}
\item $p_{ik}^j = 1$ if both $p_{ik} = 1$ and $p_{jk} = 1$ and 0 otherwise, and
\item $\bar{p}_{ik}^j = 1$ if $p_{ik} = 1$ and $p_{jk} = 0$ and 0 otherwise. 
\end{itemize}
Since country $n$ is already making product $j$, it has all the necessary capabilities for it. Hence, the probability that country $n$ starts making product $i$ can be expressed as:
$$
P(J_{n,i} = 1 |M_{n,j} = 1) = q ^ {| \bar{p}_i^j | - c_n \cdot \bar{p}_i^j } 
$$
where $q$ is the mean probability of acquiring a capability under a binomial model. We can apply Bayes' Rule:
\begin{eqnarray*}
P(M_{n,j} = 1 |J_{n,i} = 1) &=& \frac{P(J_{n,i} = 1 |M_{n,j} = 1)  P(M_{n,j} = 1)}{P(J_{n,i} = 1)} \\
& = & \frac{q ^ {| \bar{p}_i^j | - c_n \cdot \bar{p}_i^j }}{q ^ {| p_i | - c_n \cdot p_i }}P(M_{n,j} = 1) \\
& = & q ^ {-(| p_i^j | - c_n \cdot p_i^j )}P(M_{n,j} = 1)
\end{eqnarray*}
and take logarithms: 
$$
\log\left(\frac{P(M_{n,j} = 1 |J_{n,i} = 1)}{P(M_{n,j} = 1)}\right)  = (| p_i^j | - c_n \cdot p_i^j )\log(1/q).
$$
If we assume that the probability of a country having each capability is $w$, this expression becomes
\begin{equation}
E_{i,j}=\log\left(\frac{P(M_{n,j} = 1 |J_{n,i} = 1)}{P(M_{n,j} = 1)}\right)  = -| p_i^j | (1-w)\log(q).
\label{eco}
\end{equation}

%%%%%%%%%%%%%%%%%%%%%%%%%%%%%%%%%%%%%%%%%%%%%%%%%%%%%%
\subsection*{Algorithm}

We construct the ecosystem matrix $\hat{E}$ using export data from the Standard International Trade Classification (SITC) revision 4 at the 4-digit level beginning in 1984. 

In order to estimate the matrices $M$ and $J$, we measure product presences and appearances via international export competitiveness. In particular, we measure the intensity with which a country exports each product by computing its Revealed per-Capita Comparative Advantage (RpCA), which is a variant of revealed comparative advantage measure of Balassa \cite{balassa1965trade} {\it adjusting for population}. 
The RpCA that a country has in a product is defined as the ratio between the share of total exports that the product represents in the country's export basket and the share of global population. Congruently, we can also think of RpCA as the per capita export of the country in the product divided by the total per capita export in the world.
A product is over-represented in a country's export basket if its RpCA is above a threshold. 

Formally, if $X_{n,i}$ is equal to the export of country $n$ in product $i$ and $\text{pop}_n$ is the population of country $n$, then the RpCA of country $n$ in product $i$ is defined as:
\begin{equation}
R_{n,i}=\frac{X_{n,i}/\sum_{k} \text{X}_{k,i}}{\text{pop}_{n} / \sum_{k}  \text{pop}_{k}}
\end{equation}
\begin{equation}
M_{n,i} =
  \begin{cases}
    1       & \quad \text{if } R_{n,i}>\tau_1 \text{ in }t\\
    0  & \quad \text{otherwise }\\
  \end{cases}
\label{defpres}
\end{equation}
An appearance of product $i$ in country $n$ is defined as: 
\begin{equation}
\label{eqn_J}
J_{n,i} =
  \begin{cases}
    1       & \quad \text{if } R_{n,i}<\tau_0\text{ in }t_0\text{ and }R_{n,i}>\tau_1\text{ in }t_1\\
    0  & \quad \text{otherwise }
  \end{cases}
\end{equation}

Hence, we compute the entries $\hat{E}_{i,j}$ as follows: 
\begin{equation}
\label{ecoeqn}
\hat{E}_{i,j} = \log \Big( \frac{\sum_{n\in \mathcal{K}_i} M_{n,j}}{|\mathcal{K}_i|} \Big/ \frac{\sum_n M_{n,j}}{N} \Big)
\end{equation}
where $\mathcal{K}_i=\{n | J_{n,i} =1 \}$ is the set of all countries where $i$ appeared at $t_1$, and $M$ corresponds to presences measured at $t_0$.

In order to create a single composite ecosystem matrix (using data 1984-2008), we aggregate the data as follows. 
Looping over each product:
\begin{enumerate}

\item For a product $i$: we search for the set of countries $\mathcal{K}_i$ and the years (in the interval $[1989,2008]$) in which it appeared. A product appeared in a country at time $t_k$ if it was absent at $t_{k-5}$ (i.e., $R_{i,c}<\tau_0$) and present at $t_{k}$ (the mean RCA for years $t_{k}$, $t_{k+1}$ and $t_{k+2}$ is greater than $\tau_1$).

\item We then identify the products $j$ that were present in countries $K_i$ in the years preceding the product appearance at $t_k$: a product $j$ was present at  $t_{k-5}$ if the mean RCA for years $t_{k-5}$, $t_{k-4}$ and $t_{k-3}$ is greater than $\tau_1$. 

\item For each $i$ and $j$, we compute the total number of presences of each product $j$ (given an appearance of product $i$), and divide it by the number of appearance countries (e.g. the size of set $\mathcal{K}_i$).

\item We then compute the number of presences of product $j$ at $t_k$ in the set of all countries (e.g., repeat step 3 for all countries), and divide it by the total number of countries. 
 
\item Finally, the ecosystem is a log of the ratio of product presences given an appearance (step 3) and product presences overall (step 4).

\end{enumerate}

Notes: 
\begin{itemize}

\item Unless otherwise specified, following Hausmann \emph{et al} \cite{Hausmann2014implied} we set standard values for parameters for absence and presence: $\tau_0=0.05$ and $\tau_1=0.25$. 

\item The full SITC database has 185 countries in 1984. Following Hausmann \emph{et al} \cite{Hausmann2014implied}, we restrict our sample to countries with population greater than 1.2 million and total exports of at least \$1 billion in 2008. We also remove Iraq (which has severe quality issues) and Serbia-Montenegro, which split into two countries during the period studied. The sample reduces to 109 countries for the ecosystem matrix computation.

\item The full SITC database has 786 4-digit products in 1984. We omit 6 products with one-digit code '9' ('Commodities and transactions not classified elsewhere in the SITC'), and drop to 780 products.

\item We discard products if less than or equal to 5 appearances are observed during the period 1989-2008 in an effort to reduce error in the ecosystem metric. 

\item We end up with 674 products at the 4-digit level, and 63 when aggregated to the 2-digit level. 

\end{itemize}

%%%%%%%%%%%%%%%%%%%%%%%%%%%%%%%%%%%%%%%%%%%%%%%%%%%%%

\subsection*{Predicting Product Appearances}

A country $n$ has capabilities $c_n$, and products $j \in \mathcal{J}_n$. We want to compute the probability of that country $n$ will acquire the missing capabilities for the appearance of product $i$:
$$
P(J_{n,i} = 1) = q ^ {| p_i | - c_n \cdot p_i }
$$
We do not know which capabilities the country $n$ already has, but we can proxy it by products already present in the country:
$$
c_n = \mathbbm{1} \left(\sum\limits_{j \in \mathcal{J}_n} p_{j} \right)
$$
where the function $\mathbbm{1}$ takes the elements of a vector to be 1 if the elements are greater than or equal to 1. I.e., the entry $j$ of $c_n$ is 1 if at least one product that country $n$ is present in has capability $j$. 

Now we will assume that the length overlap between the products are uniformly distributed, i.e., $|p_i^j|$ is uniformly distributed. We have a measure for $|p_i^j|$ but we do not know $|p_i|$. Then Maximum Likelihood estimator up to a multiplicative factor is:
$$
\widehat{|p_i|} = \max\limits_{j} |p_i^j|
$$
Hence, we estimate the number of capabilities needed for $i$ by computing its maximum overlap with all other products $j$. 

The maximum likelihood estimator (up to the same multiplicative factor as above) for the overlap $p_i.c_n$ is then:
$$
\widehat{p_i.c_n} = \max_{j\in \mathcal{J}_n} |p_i^j|
$$
This is the maximum overlap between product $i$ and any product $j$ which is present in country $n$. Hence we estimate the total capabilities required via the overlapping capabilities between product $i$ and all countries, and subtract those already shared by country $n$.

Empirically, we estimate $\widehat{|p_i|}$ as $\max_j \hat{E}_{i,j}$ and $\widehat{p_i.c_n}$ as $\max_{j\in \mathcal{J}_n} \hat{E}_{i,j}$. Therefore, we estimate the likelihood of an appearance in the  of product $i$ in country $c$ as
$$
d^E_{i,n}=q^{\max_j \hat{E}_{i,j} -\max_{j\in \mathcal{J}_c} \hat{E}_{i,j}}
$$
where $q$ is the probability of acquiring a new capability, and $\mathcal{J}_n$ is the set of products $j$ for which country $n$ is present. In practice, in order to reduce noise, we take the mean value over the top 3 entries for each $\max_j \hat{E}_{i,j}$ and $\max_{j\in \mathcal{J}_n} \hat{E}_{i,j}$. 

%%%%%%%%%%%%%%%%%%%%%%%%%%%%%%%%%%%%%%%%%%%%%%%%%%%%%

\subsection*{The Product Space}

The Product Space \cite{Hausmann2007} is a network that models the process of industrial diversification of nations, where nodes represent products, and edges broadly represent overlap. The Product Space is built from a cross-section of data - as opposed to the time-series data required to build the Eco Space. The edge weight between two nodes is estimated using a measure of co-export - i.e., a pair of products is connected by an edge if they are exported by a similar set of countries. It has been shown that the Product Space is a good predictor of product appearances \cite{Hausmann2007, Hausmann2014implied}.

The Pearson correlation is used to infer similarity of export \cite{Hausmann2014implied}, with adjacency matrix $P$ such that
$$
P_{i,j}=\text{corr}(R_i, R_j)
$$
where $R_i$ is the vector of RpCA values for product $i$ (e.g., columns of $R_{n,i}$ defined above). The logic behind this approximation is that if a pair of products is co-exported by a large subset of countries, then these products must require a similar capability base. 

Consequently, countries are expected to move into sectors which are 'close' or similar to activities they are already successful at. From a network perspective, this is equivalent to saying that the probability of a product appearance in the future is dependent on the RpCA that the country currently enjoys in neighbouring products. Mathematically, we write the Product Space density of product $i$ in country $n$ as 
\begin{equation}
\label{eq_dens}
d^P_{i,n} = \frac{ \sum_{j} P_{j,i} R_{n,j} }{\sum_j P_{j,i}}.
\end{equation}
where the matrix $P$ represents the network proximity or adjacency matrix for the Product Space as defined above.

%%%%%%%%%%%%%%%%%%%%%%%%%%%%%%%%%%%%%%%%%%%%%%%%%%%%

\subsection*{Probit Model}

We perform a standard Probit regression for the probability of a product appearance of the form:
\begin{equation}
J_{n,i} = \Phi( \alpha +\beta_E d^E_{n,i} + \beta_P d^P_{n,i} + \gamma_i + \eta_n)
\end{equation}
where the binary variable $J_{n,i}$ is defined by Equation \eqref{eqn_J}, $\Phi$ is a normal cumulative distribution function, $d^E$ corresponds to the Eco Space density, and $d^P$ corresponds to the Product Space density, and $\gamma_i$ and $\eta_n$ are product and country fixed effects respectively. 

We construct the ecosystem for years 1984-2008, and use RpCA values from the year 2009, to compute the density metrics as given by \Cref{eq_dens} for both the Eco Space and the Product Space. Our dependent variable is defined for appearances during the 6-year period 2009-2015. Note: we condition on the product being absent at the start of the period, e.g., we only include country-product pairs that were absent in 2009.

In order to quantify the predictive power of each density metric, and their combination, we compute the AUC or Area Under the Curve of the ROC (Receiver Operating Characteristic). The ROC curve plots the rate of true positives of a continuous prediction criterion as a function of the rate of false positives. The area under the curve (AUC) statistic is equivalent to the Mann-Whitney statistic (the probability of ranking a true positive ahead of a false positive in a prediction criterion). By definition, a random prediction will find true positives and false positives at the same rate, and hence will result in an AUC = 0.5. A perfect prediction, on the other hand, will find all true positives before giving any false positive, resulting in an AUC = 1.
%
%%%%%%%%%%%%%%%%%%%%%%%%%%%%%%%%%%%%%%%%%%%%%%%%%%%

\clearpage
\begin{footnotesize}
\bibliographystyle{naturemag}

\input{NHB27.bbl}
\end{footnotesize}

\end{multicols}

\end{document}

%% file: indeg_new.tex
Agricultural, horticultural \& forestry machinery \\
Mineral tars \& their distillation products   \\
Type-setting \& founding machinery   \\
Fatty acids, oils from animal or vegetable waxes   \\
Motor vehicles for transport of ten or more persons \\
Machinery \& specialised machinery \\
Lubricating oils from petrol \& bitumin  \\
Organic \& inorganic compounds \\
Corrugated paper \\
Leather articles used in machines \& appliances   \\
Whole bovine skin leather \\
Wood and resin based chemical products \\
Strand wire, iron, steel, copper, aluminium \\
Electric, laser or other light or photon beam  \\
Silver work \\

%% file: outdeg_new.tex
Worn clothing \& other worn textile articles/rags \\
Synthetic filament \\
Machinery parts not containing electric connectors\\
Calculating machines, postage-franking \\
Hand tools, pneumatic or motor \& parts \\
Radio-broadcast receivers \\
Birds' eggs, and egg yolks \\
Plywood sheets not over 6 mm thick \\
Concrete pumps \\
Acyclic alcohols \& halogenated derivatives \\
Electric filament or discharge lamps, lamps \& parts \\
Whey, products of natural milk constituents \\
Fans \& cooker hoods incorporating a fan \\
Malt, whether or not roasted (including malt flour) \\
Works trucks, tractors, \& parts \\

%% file: regs_rpop_eco0_1.tex
\begin{tabular}{lccccccc} \hline
 & (1) & (2) & (3) & (4) & (5) & (6) & (7) \\
VARIABLES &  &  &  &  &  &  &  \\ \hline
 &  &  &  &  &  &  &  \\
Eco Density & 4.581*** & 4.540*** & 5.828*** & 4.231*** &  & 4.309*** & 3.972*** \\
 & (0.196) & (0.265) & (0.275) & (0.568) &  & (0.208) & (0.569) \\
PS Density &  &  &  &  & 1.505*** & 0.434*** & 1.762*** \\
 &  &  &  &  & (0.0952) & (0.116) & (0.262) \\
Constant & -5.863*** & -5.397*** & -6.095*** & -4.298*** & -2.332*** & -5.698*** & -4.117*** \\
 & (0.164) & (0.234) & (0.283) & (0.469) & (0.0197) & (0.169) & (0.469) \\
 &  &  &  &  &  &  &  \\
Observations & 46,919 & 39,193 & 27,178 & 22,748 & 47,576 & 46,919 & 22,748 \\
Country FE & No & Yes & No & Yes & No & No & Yes \\
Product FE & No & No & Yes & Yes & No & No & Yes \\
Pseudo R2 & .072 & .115 & .128 & .178 & .022 & .074 & .184 \\
 AUC & .736 & .789 & .792 & .830 & .664 & .738 & .836 \\ \hline
\multicolumn{8}{c}{ Robust standard errors in parentheses} \\
\multicolumn{8}{c}{ *** p$<$0.01, ** p$<$0.05, * p$<$0.1} \\
\end{tabular}